\documentclass[aip,apl,reprint,twocolumn]{revtex4-1}
\usepackage{lipsum}% http://ctan.org/pkg/lipsum
\usepackage{graphicx} 
\usepackage{bm} 
\usepackage{dcolumn}
\usepackage{amssymb,amsmath}
\bibstyle{apsrev}
\usepackage{hyperref}
\usepackage{color}
\usepackage[normalem]{ulem}
\usepackage{bm}

\usepackage{siunitx}

\begin{document}

\title{Non-reciprocity of spin waves in magnetic nanotubes with helical equilibrium magnetization}

\author{M.M. Salazar-Cardona}
\affiliation{Departamento de F\'isica, Universidad Cat\'olica del Norte, Av. Angamos 0610, Antofagasta, Chile.}
\author{L. K\"orber}
\affiliation{Helmholtz-Zentrum Dresden-Rossendorf, Institute of Ion Beam Physics and Materials Research,
Bautzner Landstr. 400, 01328 Dresden, Germany}
\affiliation{Fakult\"at Physik, Technische Universit\"at Dresden, D-01062 Dresden, Germany}
\author{H. Schultheiss}
\affiliation{Helmholtz-Zentrum Dresden-Rossendorf, Institute of Ion Beam Physics and Materials Research,
Bautzner Landstr. 400, 01328 Dresden, Germany}
\affiliation{Fakult\"at Physik, Technische Universit\"at Dresden, D-01062 Dresden, Germany}
\author{K.  Lenz}
\affiliation{Helmholtz-Zentrum Dresden-Rossendorf, Institute of Ion Beam Physics and Materials Research,
Bautzner Landstr. 400, 01328 Dresden, Germany}
\author{A. Thomas}
\affiliation{Institute for Metallic Materials, Leibniz Institute of Solid State and Materials Science, 01069 Dresden, Germany}
\affiliation{Technische Universit\"at Dresden, Institut f\"ur Festk\"orper- und Materialphysik, 01062 Dresden, Germany}
\author{K. Nielsch}
\affiliation{Institute for Metallic Materials, Leibniz Institute of Solid State and Materials Science, 01069 Dresden, Germany}
\affiliation{Technische Universit\"at Dresden, Institute of Materials Science, 01062 Dresden, Germany}
\author{A. K\'akay}
\affiliation{Helmholtz-Zentrum Dresden-Rossendorf, Institute of Ion Beam Physics and Materials Research,
Bautzner Landstr. 400, 01328 Dresden, Germany}
\author{J. A. Ot\'alora}
\email[Corresponding author: ]{jorge.otalora@ucn.cl}
\affiliation{Departamento de F\'isica, Universidad Cat\'olica del Norte, Av. Angamos 0610, Antofagasta, Chile.}

\date{\today}
\begin{abstract}
% the allowed words for the APL abstract is 250, currently we have 216
Spin waves (SWs) in magnetic nanotubes have shown interesting nonreciprocal properties in their dispersion relation, group velocity, frequency linewidth and attenuation lengths. The reported chiral effects are similar to those induced by the Dzyaloshinskii-Moriya interaction, but originating from the dipole-dipole interaction. Here we show, that the isotropic-exchange interaction can also induce chiral effects in the SW transport; the so-called Berry phase of SWs. We demonstrate that with the application of magnetic fields, the nonreciprocity of the different SW modes can be tuned between the fully dipolar governed and the fully exchange governed cases, as they are directly related to the underlaying equilibrium state. In the helical state, due to the combined action of the two effects every single sign combination of the azimuthal and axial wave vectors leads to different dispersion, allowing for a very sophisticated tuning of the SW transport. A disentanglement of the dipole-dipole and exchange contributions so far was not reported for the SW transport in nanotubes. Furthermore, we propose a device based on coplanar waveguides that would allow to selectively measure the exchange or dipole induced SW nonreciprocities. In the context of magnonic applications, our results might encourage further developments in the emerging field of 3D magnonic devices using curved magnetic membranes.

\end{abstract}
\pagestyle{plain} 
\maketitle

Magnonics, the field that harnesses spin waves (SWs) as information carriers, promises an unprecedented augment of capabilities for information transfer, processing and sensing. Applications employing magnons have been highlighted by their lower power consumption, reconfigurable multi-functionality, faster operation rates, and by its potential integrability to spintronic and electronic environments via the interconversion mechanisms among magnons, spin accumulation and electric charge. \cite{TserkovnyakPRL02,KajiwaraNATURE010,MadamiNATNANO11,SandwegPRL11,DemidovNATMAT12} 
The magnon transistor for all-magnon data processing,\cite{ChumakNCOM14} SW multiplexers,\cite{VogtNCOM14} magnonic beam splitter,\cite{SadovnikovAPL15} magnonic diodes,\cite{LanPRX15} SW logic gates\cite{KlinglerAPL14,IEEE15Radu,khitunJAP11} and phase-to-amplitude/amplitude-to-phase SW converters\cite{BracherSCREP16,TockARXIV18,TalmelliARXIV18} are just a few illustrative examples.

Magnonic applications are build mainly on two-dimensional (2D) planar layouts. However, when the geometry is extended to three dimensional (3D) curved magnetic membranes, curvature-induced magnetochiral effects emerge expanding the toolbox of SW control, 
 which can be exploited for further enhancements of magnonic devise capabilities.\cite{HertelSPIN13,GaidideiPRL14,OtaloraPRL16,StreubelJPDAP16,OtaloraPRB17,FernandezNATCOM17,StanoHMM18,ShekaPRB17,GaidideiJPAMT17,OtaloraPRB18,ROADMAP20,MakarovCOMPHYS2020} Recently has been predicted that the surface curvature leads to the renormalization of the interactions and as a consequence will affect both the static configurations and the magnetization dynamics, including SW transport.\cite{OtaloraJAP15,OtaloraPRL16,OtaloraPRB17,OtaloraPRB18,MakarovCOMPHYS2020} 

In this Letter, we discuss the curvature induced asymmetric SW dispersion of dipole-dipole and isotropic-exchange origin in nanotubes with thin magnetic shell. The latter is referred to as the Berry phase of SWs\cite{DugaevPRB04}. The strength of the individual contributions to the non-reciprocal SW transport can be tuned by changing the underlying equilibrium state: (i) for a vortex state purely diplolar effects occur, (ii) in the axially magnetized state purely exchange-driven effects occur, whereas (iii) a combination of both governs the SW transport in the helical state. We propose a magnonic transducer based on coplanar waveguides that would allow to measure experimentally and quantify the exchange and dipolar contributions to the asymmetric SW dispersion. The theory for the expected scattering parameters by the suggested coplanar waveguide setup is also presented.

\begin{figure}[t]
    \centering
    \includegraphics[width=8.5cm]{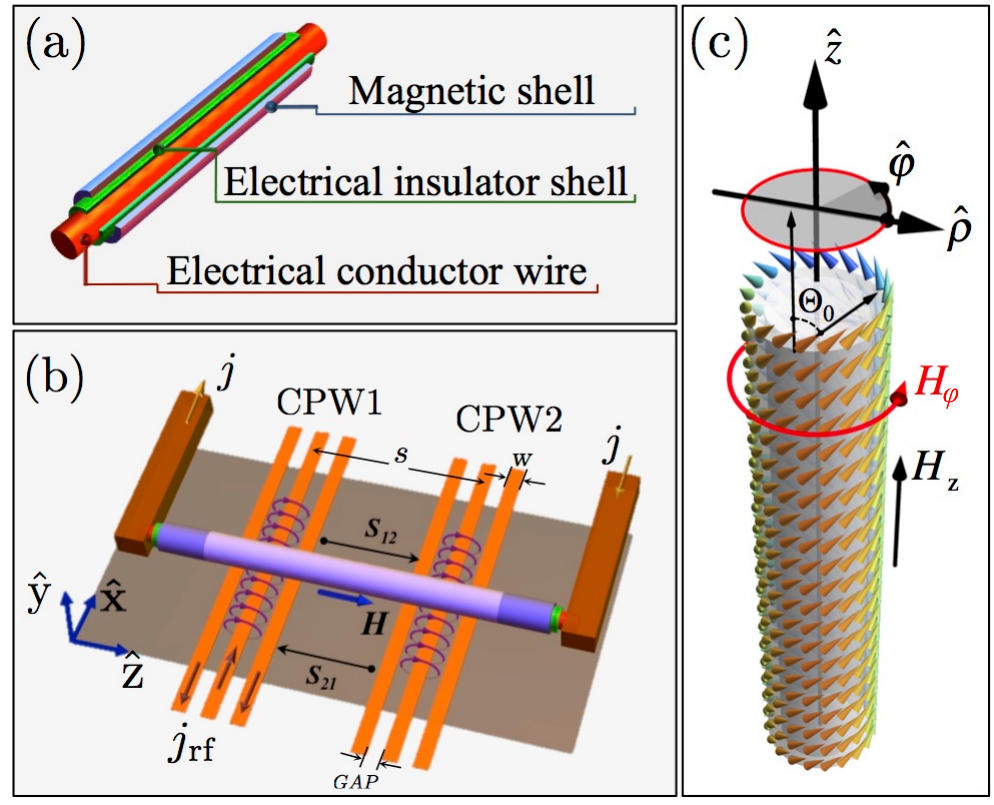}
    \caption{(a) Magnonic component comprised of core-shell MNT made of an internal conductor wire, electrical insulator layer and outer magnetic shell. (b) The MNT is placed on top of two identical and parallel coplanar wave guides, CPW1 and CPW2. The ground and signal lines are separated by a gap $GAP$. The geometrical dimensions of the CPWs are $\textsl{w}=\text{GAP}=250$ nm and a separation of $s=5$ $\mu$m between the two CPW centers. The internal conductor wire is connected to two electrical terminals for the injection of a DC electric current $j$, which creates a circular magnetic field $H_\varphi$ in the magnetic shell. The signal will be measured via the transmission coefficients $S_{12}$ and $S_{21}$. 
c) The nanotube in a helical equilibrium configuration described by the angle $\Theta_0$.\cite{OtaloraJAP15,SalinasSREP18}}
    \label{fig:fig1}
\end{figure}

The focus of our study is a \textit{permalloy} magnetic nanotube (MNT) with the following geometrical dimensions and material parameters, outer(inner) radius $R=105$ nm ($r=95$ nm), length $L=15$ $\mu$m, saturation magnetization $\mu_0M_s=1$ T, exchange stiffness constant $A=13$ pJ/m and exchange length $l_{ex}=5.8$ nm. The nanotube size is chosen based on the recent progress in synthesis of MNTs with similar sizes, low damping and high conductivity~\cite{GrundlerACSAMI2020}. The equilibrium state for these material parameters and geometrical dimension in the absence of any external fields is homogeneously magnetized state along the long axis with two oppositely curling vortex caps at the ends~\cite{LanderosPRB09,WyssPRB17,StanoSciPP18,ZimmermannNANOLETT18}. The vortex state can be achieved with an azimuthal field. The critical field to stabilize the circular magnetic state in an infinitely long MNT is $\mu_0H_{\text{crit}}\approx 3.4$ mT~\cite{OtaloraJAP15}. Inducing this field value requires the injection of a minimum electric current density of $j\sim 10^{11}$ A/m$^2$ into the electrical conductor [Fig.~\ref{fig:fig1}(b)].~\cite{OtaloraJAP15} For the current study, in order to gradually tune the equilibrium magnetic state between the axially saturated to the vortex state through a helical state we will use a combination of a circular and axial fields.

Following, we present a theory for the linear SW properties in a MNT with helical equilibrium magnetization. We will consider exchange interaction, dipolar interaction and uniaxial magnetocrystalline anisotropy.
The equilibrium magnetization of the nanotube is written as 
\begin{equation}
    \bm{M}_0(\rho,\varphi) = M_\mathrm{s}\big[\sin(\Theta_0) \hat{\bm{\varphi} }+\cos(\Theta_0) \hat{\bm{z}}\big]
\end{equation}
with the saturation magnetization $M_\mathrm{s}$ and $\Theta_0$ denoting the (spatially independent) angle of the magnetization with the $z$ axis [Fig.~\ref{fig:fig1}(c)]. This state can be stabilized using external fields in $\varphi$ and $z$ direction ($H_\varphi$ and $H_z$, respectively), and a uniaxial anisotropy along the $z$ axis with negative constant $K\leq 0$. In the limiting cases of $\Theta_0=\pi/2$ and $\Theta_0=0,\pi$, the helical state is denoted as vortex- and axial state, respectively.

In particular, the angle $\Theta_0$ can be found from the equilibrium condition
\begin{equation}\label{eq:equilibirum-condition}
\begin{split}
        &\big[H_\varphi - ({H}_\mathrm{ex} + {H}_\mathrm{uni})\sin(\Theta_0)\big]\cos(\Theta_0)\\
        &\qquad\qquad\qquad\qquad\qquad\qquad - H_z \sin(\Theta_0) = 0,
\end{split}
\end{equation}
Here, the ${H}_\mathrm{uni}=2K/(\mu_0 M_\mathrm{s})$ is the maximum ($K>0$) or minimum ($K<0$) of the anisotropy field strength, and ${H}_\mathrm{ex}=M_\mathrm{s}l_\mathrm{ex}^2/b^2$ is the maximum exchange field magnitude for the magnetization in the vortex state ($\Theta_0=\pi/2$).\cite{OtaloraPRL16,OtaloraPRB17} The factor $b^{-2}$ is  obtained from the radial average of inverse second power of the radial coordinate, as $ b^{-2}\equiv(2\pi/S)\int_{r}^{R} \rho^{-1}d\rho= 2\pi \ln(R/r)/S $
where $r$ and $R$ are the inner and outer radii of the nanotube respectively, and $S = \pi(R^2 - r^2)$ is the cross-section area of the nanotube. From Eq.~\eqref{eq:equilibirum-condition} it follows that the angle $\Theta_0 \equiv \Theta_0(H_\varphi,H_z)$ implicitly depends on the two external fields, and, furthermore, that any state with $\Theta_0 \neq 0, \pi$ (any other state than the axial state) can only be stable if the azimuthal external field
\begin{equation}
    \vert H_\varphi \vert \geq H_\mathrm{crit} = H_\mathrm{ex} + H_\mathrm{uni}.
\end{equation}
The critical field $H_\mathrm{crit}$ can be reduced to zero using a (negative) uniaxial anisotropy, as for example shown for hexagonal tubes in \textcite{ZimmermannNANOLETT18}, that is achieved by an incident angle deposition. 
 
Let us focus now on the dynamical magnetic response of the nanotube. In general, the SWs in a thin-shell MNT can be categorized by an azimuthal mode number $n$ and the wave vector $k_z$ in $z$-direction, where the index $n$ takes positive and negative values and counts the periods of the waves along the $\varphi$ direction. The SW dispersion relation $\omega_n(k_z)$ and dynamic susceptibility in the vortex state (at $H_z=0$ and $K=0$) have been studied previously.\cite{OtaloraPRL16,OtaloraPRB17,OtaloraPRB18} In these works, it was shown that the asymmetric SW dispersion in $z$ direction is the consequence of the curvature-induced broken mirror symmetry of the dynamic volume charges. Moreover, modes with the same $k_z$ but opposite azimuthal index, $\pm n$, are degenerate, hence the SWs form azimuthal standing waves with $2|n|$ nodal lines. With the application of an axial magnetic field $H_z\neq 0$, the frequency degeneration regarding the sign of $n$ is broken, consequently the formation of azimuthal standing waves is disrupted and conditions to observe individual azimuthal modes on the nanotube mantle separately is therefore possible, for instance with the use of CPWs as we propose here. Similar to Refs.~\citenum{OtaloraPRL16,OtaloraPRB17}, the dispersion relation $\omega_n(k_z)$ is obtained by setting the Gilbert damping to cero, solving the linearized equation of motion of magnetization and can be written in the from
\begin{equation}
    \frac{\omega_n (k_z)}{\omega_M} =\mathcal{A}_n(k_z) + \sqrt{\mathcal{B}_n(k_z)\mathcal{C}_n(k_z)}
\end{equation}
where $\omega_M = \gamma \mu_0 M_\mathrm{s}$ is the characteristic frequency and $\gamma$ is the gyromagnetic ratio of the material at hand. The symbols $\mathcal{A}_n(k_z)$, $\mathcal{B}_n(k_z)$ and $\mathcal{C}_n(k_z)$ denote the normalized dynamic stiffness fields (detailed expressions are found in the supplementary material). All three depend on the angle $\Theta_0$ and therefore on the external fields $H_\varphi$ and $H_z$, whereas $\mathcal{B}_n$ and $\mathcal{C}_n$ constitute the stiffness fields along the $\rho$ and $\varphi$ directions, respectively.\cite{OtaloraPRB18} The only stiffness field which can create a dispersion asymmetry in the sense 
\begin{equation}
   \Delta\omega_n(k_z) \equiv \omega_n (k_z) - \omega_{-n} (-k_z) \neq 0
\end{equation}
is the normalized magnetochiral field $\mathcal{A}_n(k_z)$, since the fields $\mathcal{B}_n$ and  $\mathcal{C}_n$ are symmetrical in $n$ and $k_z$. A result, the dispersion asymmetry is directly given by $\Delta\omega_n(k_z) = 2\omega_M \mathcal{A}_n(k_z)$. The magnetochiral field is given by
\begin{equation}\label{eq:A}
\begin{split}
        \mathcal{A}_n(k_z) &= \mathcal{K}_n(k_z) \sin(\Theta_0) \\
        & \qquad - \big[\mathcal{N}_n(k_z) - 2(H_\mathrm{ex}/M_\mathrm{s}) n\big]\cos(\Theta_0).
\end{split}
\end{equation}

Terms $\mathcal{K}_n$ and $\mathcal{N}_n$ are hyper-geometrical functions which depend on the inner and outer radii of the nanutube, $r$ and $R$ (see supplementary material). They are of purely dipolar nature and appear when expanding the dipolar magnetostatic potential of the individual SW modes in terms of cylindrical-harmonics basis functions as
\begin{align}
    \mathcal{K}_n(k_z)&=\frac{\pi }{S}\int _0^{\infty}dq\frac{q^2 k_z}{q^2+k_z^2}\Gamma _n[q]\Lambda_n[q]\\
    \mathcal{N}_n(k_z)&=n\frac{\pi }{S}\int _0^{\infty }dq\frac{q^2}{q^2+k_z^2}\Gamma _n[q]I_n[q].
\end{align}
where $I_n(q)=\int _r^Rd\rho J_n (q \rho)$, $\Lambda_n (q)=\int _r^Rd\rho \ \rho J_n (q \rho)$, and $\Gamma _n (q)=\Lambda_{n-1}(q)-\Lambda_{n+1}(q)$. Here, $J_n(x)$ is the Bessel functions of first kind and order $n$. 

We can see from Eq.~\eqref{eq:A} that for the vortex-state, $\Theta_0=\pi/2$, the dispersion asymmetry is solely determined by the dipolar interaction $\Delta \omega_n(k_z) = 2\omega_M \mathcal{K}_n(k_z)$, as was reported in Ref.~\citenum{OtaloraPRL16}. Let us again note that $\mathcal{K}_n(k_z)$ is a purely geometrical quantity while $\omega_M$ is a material parameter. On the other side, in the axial state, $\Theta_0 = 0,\pi$, the asymmetry is comprised of both dipolar and exchange interaction $\Delta \omega_n(k_z) = 2\omega_M \big[ \mathcal{N}_n(k_z) -2 (H_\mathrm{ex}/M_\mathrm{s}) n\big]$, while for $k_z=0$ it is purely exchange dominated. In fact, in this case, modes propagating in opposite directions along the circumference ($\pm n$) accumulate a different phase, which is the exchange-induced Berry phase for SWs.\cite{DugaevPRB04} Up to now, no experimental evidence of this Berry phase is available. Thus, in the general helical state, a combination of dipolar and exchange introduced asymmetries are present. Therefore, by changing the angle $\Theta_0$ by varying the external field $H_z$, a versatile control of the SW dispersion and resulting transport in nanotubes is possible. From a material point of view, the exchange asymmetry scales with the ratio $A_\mathrm{ex}/M_\mathrm{s}$ (with the exchange stiffness constant $A_\mathrm{ex}$) whereas the dipolar asymmetry scales with $M_\mathrm{s}$ (see supplementary material). Thus, materials with large $A_\mathrm{ex}/M_\mathrm{s}$ ratio and large $M_\mathrm{s}$ are favorable. From a geometrical point of view, both asymmetries are larger for smaller outer radii $R$. Note that, however, increasing $A_\mathrm{ex}/M_\mathrm{s}$ and decreasing $R$ requires a larger azimuthal field $H_\varphi$ or a larger (negative) anisotropy $K$ to stabilize the helical state.

 %+++++++++ Figure2:  +++++++++++
\begin{figure}[t!]
\begin{center}
\includegraphics[width=5.5cm]{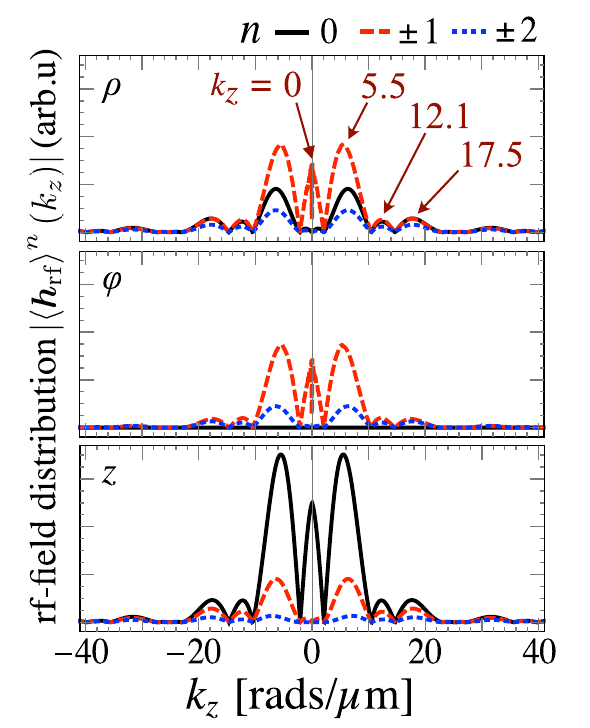}
\end{center}
\caption{(Color online). Wave vector distribution of the magnetic fields $H_{\rho}[\vec r]$, $H_{\varphi}[\vec r]$ and $H_z[\vec r]$ created by a CPW with geometrical dimensions $\textsl{w}=\text{GAP}=250$ nm and a separation of $s=5$ $\mu$m between the signal and ground lines. Here $(\rho,\varphi,z)$ are the cylindrical coordinates. Representative peaks of the vector distribution are mostly given at $n=0,\pm1,\pm2$ and $k_z$ with values of $|k_z|\in \{0, 5.5, 12.1, 17.5\}$ rads/$\mu$m.}
\label{CPWSpectrum}
\end{figure}
%+++++++++++++++++++++++++++++++++++

In the following, we would like to discuss how to disentangle the two curvature-induced contributions in an CPW experiment and showcase the evolution of the asymmetry and thus the tunability of the dispersion with the angle $\Theta_0$. For simplicity, we consider the case $K=0$, thus no magnetocrystalline anisotropy, in which case the critical field $\mu_0 H_\mathrm{crit}=\SI{3.5}{\milli\tesla}$ as mentioned above. Let us assume that SWs are excited in the nanotube using the CPW2 [Fig.~\ref{fig:fig1}(b)]. The rf-field distribution produced by a CPW in the Fourier space $\{n,k_z\}$ is shown in Figure \ref{CPWSpectrum}. A main maximum peak is observed at $k_z=5.5$ rads/$\mu$m and $n=\pm1$, and with a width of $\Delta k_z=7.5$ rads/$\mu$m. This peak is in agreement with the literature where a maximum peak around $k_z\approx \pi/(2\textsl{w})$ is expected. \cite{DurrPHDTHESIS12,SchwarzePHDTHESIS13,VlaminckPRB10} Note, that there is also a peak at $k_z=0$ thus the Kittel modes or ferromagnetic resonance (FMR) modes can also be excited.

 %+++++++++ Figure3:  +++++++++++
\begin{figure*}[th!]
\begin{center}
\includegraphics[width=15cm]{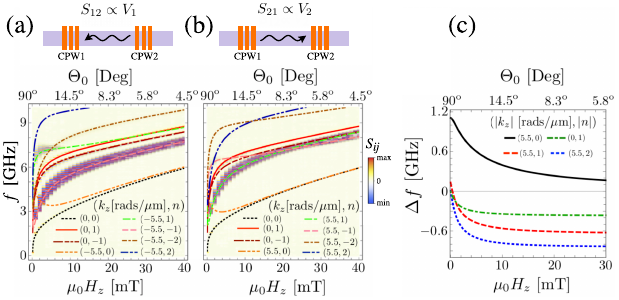}
\end{center}
\caption{(Color online). (a),(b) The scattering parameter $S_{12}$ ($S_{21}$) (color coded) induced in CPW1 (CPW2) by exciting SWs in the MNT with CPW2 (CPW1) in function of RF frequency and applied magnetic field $H_z$. The difference in the maximum values are related to the Damon-Eshbach geometry and the asymmetric frequency linewidth induced by the curvature. c) Frequency asymmetry $\Delta f$ as a function of the applied field $H_z$ for one exemplary excitation wavelength $k_z=5.5$ rads$/\mu$m and three azimuthal modes $n=0,1,2$.}
\label{CPWImpAsymImp}
\end{figure*}
%++++++++++++++++++++++++++++++++++++

The transmission coefficients or scattering parameters $S_{12}$ ($S_{21}$) related to CPW1 (CPW2), the quantity usually measured in experiments using a VNA (Vector Network Analyzer) is proportional to the induced voltage, $S_{12}\propto V_1$ and $S_{21}\propto V_2$. 
The voltages $\textsl{v}_1^{(n)}(k_z,\omega)$ induced into the CPW1 through the magnetic flux of the individual propagating SW modes are determined by the dynamic susceptibility $\Hat{\bm{\chi}}_n(k_z)$ (also called Polder tensor) as well as by the Fourier components of the generated microwave field $\bm{h}_{\mathrm{RF},n}(k_z)$ by CPW2, which means $\textsl{v}_1^{(n)} \equiv \textsl{v}_1^{(n)}(\Hat{\bm{\chi}}_n,\bm{h}_{\mathrm{RF},n},\omega)$. The full expressions for the dynamic susceptibility and induced voltages are lengthy and therefore only shown in the supplementary material. Once the voltages induced by each mode are obtained, the total induced (complex-valued) voltage into CPW1 $\Tilde{V}_1(\omega)$ is given by integrating over all wave vectors $k_z$ and summing up over all azimuthal mode numbers $n$, while the respective scattering parameter $S_{21}$ is proportional to the real part of it.
\begin{equation}\label{eq:Impedance}
S_{12}\propto \mathrm{Re}\,\Tilde{V}_1(\omega)=s\sum_{n=-\infty}^{\infty}\int_0^{\infty} \mathrm{d} k_z \ \mathrm{Re}\,\big[ \textsl{v}_1^{(n)}(k_z,\omega)\big]
\end{equation}
Here, $s$ is the spacing between the CPW. A similar expression is found for the complex voltage $\Tilde{V}_2$ measured at CPW2 induced by CPW1.

The scattering parameters calculated using Eq.~\eqref{eq:Impedance} as a function of the applied magnetic field $H_z$ and excitation frequency are shown colour coded in figure \ref{CPWImpAsymImp}. In these predictions, we can track the "measured" resonances to reciprocal and nonreciprocal SWs  by the SW eigenfrequencies $f_n(k_z)=\omega_n(k_z)/(2\pi)$ as lines in the figure. The strongest signals are consistent with the larger peaks from the rf-field distribution (see figure \ref{CPWSpectrum}). Such peaks correspond to the modes $n=0,\pm1$ and are centered at the wave vectors $k_z=0,\pm 5.5$ rads/$\mu$m. Note that the FMR mode, corresponding to the SW with $k_z=0$ and $n=0$ (black-dotted curve), can be resolved and is reciprocal as expected. Nonreciprocities can be observed by comparing the cases $\{n,k_z\}$ to $\{-n,-k_z\}$ between the spectra $S_{21}$ and $S_{12}$ in figure \ref{CPWImpAsymImp}(a,b), and we try to quantify the frequency asymmetries $\Delta f\equiv f[n,k_z]-f[-n,-k_z]$ (for the most prominent wavevectors that are resolved by the SW scattering parameters figure \ref{CPWSpectrum})) as function of $H_z$, shown in figure \ref{CPWImpAsymImp}c).

The asymmetry induced only by the exchange interaction or Berry phase of spin waves ($\Delta f_{\text{ex}}^n$) can be measured in the axial saturated state for the mode with $|n|=1$ at $k_z=0$, represented by the red line in the spectrum in panels (a,b). This is possible since at $k_z=0$ the effect from the dipolar interaction vanishes, namely at $k_z=0$ $\mathcal{K}_{k_z}^n=\mathcal{N}_{k_z}^n=0$, hence the asymmetry depends only on the dynamic exchange field $\Delta f=\Delta f_{\text{ex}}^n=\omega_{M_s}\mathcal{A}_{0}^{n}/\pi=-2\omega_{M_s}h_u n\cos(\Theta_0)/\pi$)\cite{DugaevPRB04}.

In the helical state, the dipolar-dominated asymmetries ($\Delta f_{\text{d}}^{k_z}$) could be measured, since for $k_z\neq 0$ and $n=0$ we have $\mathcal{N}_{k_z}^0=0$, hence $\Delta f=\Delta f_{\text{d}}^{k_z}=\omega_{M_s}\mathcal{A}_{k_z}^{0}/\pi=\omega_{M_s}\mathcal{K}_{k_z}^{0}/\pi$. The maximum asymmetry is obtained for the vortex state at $H_z=0$. For $n\neq 0$ and $k_z\neq 0$ both the exchange and dipolar effects contribute to the asymmetries in the helical state.

The blue dotted and red dashed curves in figure \ref{CPWImpAsymImp}(c) show the simultaneous contribution of exchange and dipolar effects in the frequency asymmetry as a function of $H_z$. For these modes the asymmetry is around or even above $|\Delta f_{\text{d}}^{5.5}|\approx0.6$ GHz for all azimuthal modes $n \neq 0$. A maximum value of the dipolar-dominated frequency shift is approximately $1.1$ GHz and decreases to $|\Delta f_{\text{d}}^{k_1}|\approx0.25$ GHz with increasing $H_z$ (black line). On the other side, the purely exchange induced asymmetry is zero at $H_z=0$ and increases with increasing field up to $|\Delta f_{\text{ex}}^{1}|\approx0.35$ GHz (see green dot-dashed curve). 

It is worth mentioning, that $|\Delta f_{\text{ex}}^{n}|$ can be increased to the GHz range by reducing the nanotubes radius. For instance, for Permalloy nanotubes with radius $R\leq65$ nm and thickness $10$ nm an asymmetry of $|\Delta f_{\text{ex}}^{1}|> 1$ GHz is obtained.  Larger order modes $|n|>1$ at $k_z=0$ can show $|n|$ times larger exchange-dominated frequency asymmetries than in $|n|=1$, nevertheless, a different CPW design than the one proposed here would be required for measuring them.

It is important to comment on the total SW wavelength ($\lambda=2\pi/k$) that could be excited/measured on the MNT using the CPWs. Indeed, our predictions show that the curvature of MNTs would allow exciting SWs with large (micrometer) and short (sub-micrometer) wavelengths $\lambda=2\pi/k$, where $k=\sqrt{k_z^2+\bar k_{\varphi}^2}$ is the total wave vector magnitude and $\bar k_{\varphi}=n/\bar \rho$ is the azimuthal wave vector. Hence, for those wavenumbers that can be resolved by the transmission coefficients, a micrometer wavelength $\lambda=1.14$ $\mu$m is obtained for $|k_z|=5.5$ rads/$\mu$m and $n=0$, whereas the shortest wavelength $\lambda=300$ nm is given for $|k_z|=5.5$  rads/$\mu$m and $|n|=2$. Notice that the wavelength can be reduced even further to $\lambda=240$ nm in the case that the mode $|n|=2$ and wave vector $|k_z|=17.5$ rads/$\mu$m can be resolved.

We have shown that the dispersion of spin waves in nanotubes is influenced by the curvature through the dipole-dipole and isotropic exchange interactions. The latter is referred to as the Berry phase of SWs\cite{DugaevPRB04}. The strength of the individual contributions to the non-reciprocal SW transport could be measured by tuning the equilibrium state with an external field from vortex to axial state. In vortex state purely dipolar effects, in axially magnetized state purely exchange effects are governing the SW transport. For the give geometrical and material parameters asymmetries in the SW dispersion at the GHz range, at sub-micrometer wavelengths and on-demand controllability of its curvature-induced nonreciprocal properties have been predicted. In the context of magnonic applications, our results might encourage further developments in the emerging field of 3D magnon devices using curved magnetic membranes.

\section*{Supplementary Material}
Detailed derivations and lengthy formulas are in the supplementary materials.

\section*{Acknowledgments}
JAO gratefully acknowledge financial support to the Fondecyt Iniciaci\'on grant No 11190184. Financial support by the Deutsche Forschungsgemeinschaft within the program KA 5069/1-1 and KA 5069/3-1 are acknowledged.

\section*{Data Availability}
The data that supports the findings of this study are available within the article [and its
supplementary material].

\section*{Author contributions}

 J.A.O. developed the analytical model and wrote the manuscript. All authors interpreted and discussed the results and co-wrote the manuscript.

\section*{Additional information}
The authors declare no competing financial interests. Reprints and permission information is available online at http://xxxxxxxxx. Correspondence and requests for materials should be addressed to J.A.O.

\section*{References}
%\bibliography{TailoringRef}
\bibliographystyle{apsrev4-1}
%%%%%%%%%%%%%%%%%%%%%%%%%%%%%%%%%%%%%%%%%%
%%%%%%%%%%%%%%%%%%%%%%%%%%%%%%%%%%%%%%%%%%
%

\end{document}